\documentclass[aps,prl,superscriptaddress,amsmath,amssymb,a4paper,twocolumn,longbibliography]{revtex4-1}
\usepackage{comment}

\usepackage[utf8]{inputenc}
\usepackage{amsmath}
\usepackage{amssymb,amsfonts,latexsym}
\usepackage{bm}
\usepackage[mathcal]{euscript}
\usepackage{graphicx}
\graphicspath{{Figures/}} 
\usepackage{epsfig}
\usepackage{color}
\usepackage{pifont,wasysym,marvosym}
\usepackage{textcomp}
\usepackage{comment}
\usepackage{epstopdf}
\usepackage{hyperref}

\usepackage{soul}

\usepackage{accents}

\usepackage{xpatch}
\usepackage{array,booktabs}
\newcolumntype{L}{@{}>{\kern\tabcolsep}c<{\kern\tabcolsep}}

\usepackage[table]{xcolor}

\begin{document}
\title{Tuning flow asymmetry with bio-inspired soft leaflets}
\begin{abstract}

In Nature, liquids often circulate in channels textured with leaflets, cilia or porous walls that deform with the flow. These soft structures are optimized to passively control flows and inspire the design of novel microfluidic and soft robotic devices. Yet so far the relationship between the geometry of the soft structures and the properties of the flow remains poorly understood. Here, taking inspiration from the lymphatic system, we devise millimetric scale fluidic channels with asymmetric soft leaflets that passively increase (reduce) the channel resistance for forward (backward) flows. Combining experiments, numerics and analytical theory, we show that tuning the geometry of the leaflets controls the flow properties of the channel through an interplay between asymmetry and nonlinearity. In particular, we find the conditions for which flow asymmetry is maximal. Our results open the way to a better characterization of biological leaflet malformations and to more accurate control of flow orientation and pumping mechanisms for microfluidics and soft robotic systems.
\end{abstract}

\author{Martin Brandenbourger}
\affiliation{Institute of Physics, Universiteit van Amsterdam, Science Park 904, 1098 XH Amsterdam, The Netherlands }
\author{Adrien Dangremont} 
\affiliation{Institute of Physics, Universiteit van Amsterdam, Science Park 904, 1098 XH Amsterdam, The Netherlands }
\author{Rudolf Sprik}
\affiliation{Institute of Physics, Universiteit van Amsterdam, Science Park 904, 1098 XH Amsterdam, The Netherlands }
\author{Corentin Coulais}
\affiliation{Institute of Physics, Universiteit van Amsterdam, Science Park 904, 1098 XH Amsterdam, The Netherlands }

\maketitle


\section*{Introduction}

Every living system is composed of soft elements containing fluids \cite{vogel2000}. In many cases, these fluids circulate within the living body to transport functional fluids such as blood or lymph via variety of pumping mechanisms   \cite{Sotiropoulos2016,Moore2018,Jaffrin1971,vogel2000}. One particularly robust mechanism to transport fluids at low Reynolds numbers takes place in the lymphatic system. Soft valves distributed along the lymphatic channel are known to promote forward flows rather than backward flows at any flow rate \cite{Moore2018}. In practice, self-regulated contractions of muscles surrounding the lymphatic channel \cite{kunert2015} induce liquid flows along the channel. The soft valves made of asymmetric soft leaflets passively deform with the liquid flow, which as a result alters the geometry of the lymphatic channel therefore increasing the flow resistance for backward flows and reducing it for forward flows. 

\begin{figure}[b!]
	\centering
\includegraphics[scale=0.85]{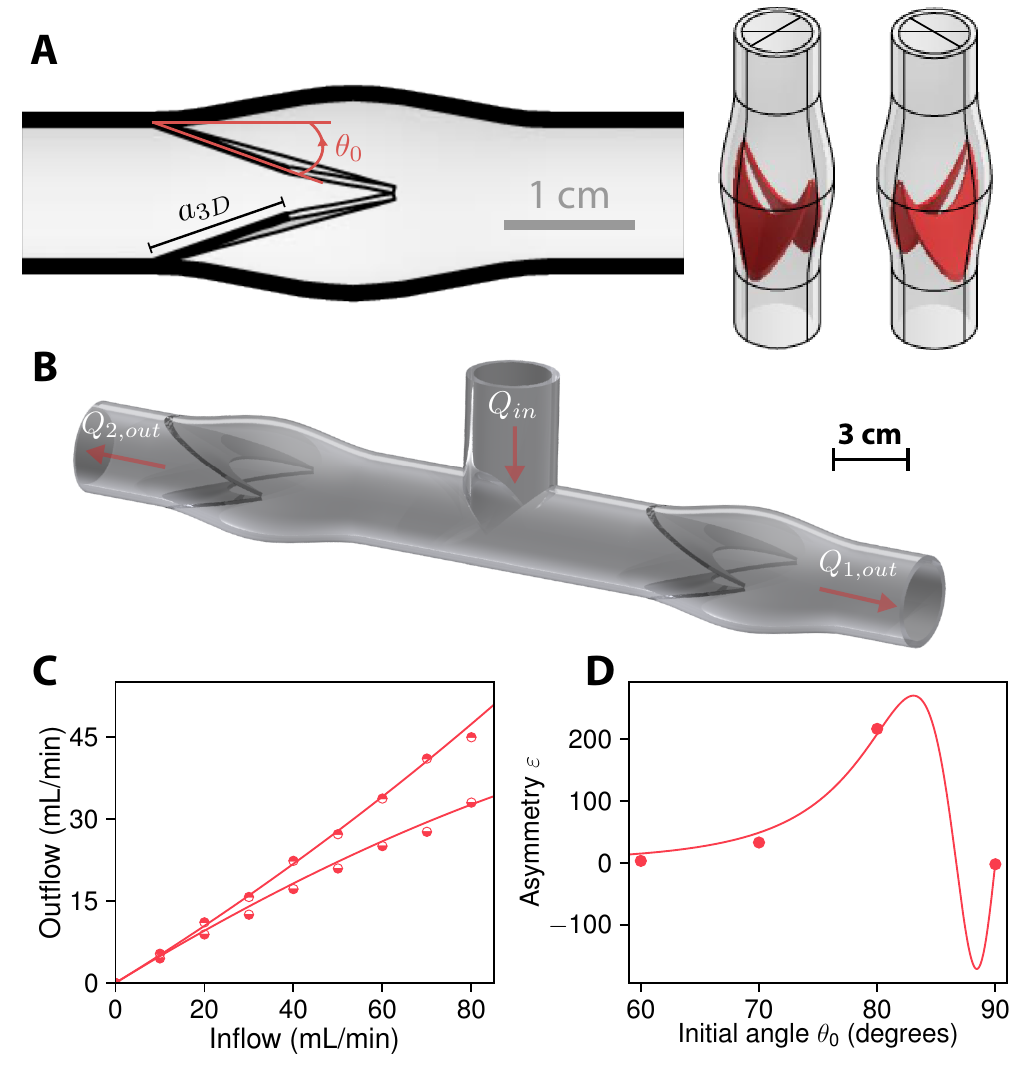}
			
	\caption{(A) Schematic of a 3D valve leaflet inspired from 3D scanners of lymphatic leaflets \cite{zawieja2009}. (B) Schematic of the setup mimicking two lymphatic valves in a T-junction geometry. (C) Outflows $Q_{out, 1}$ and $Q_{out, 2}$ as a function of the inflow $Q_{in}$ for an initial opening $\theta_0=70^\circ$. For each inflow $Q_{in}$, $Q_{out, 1}  > Q_{out, 2}$.  The red line corresponds to a fit given by Eq. (\ref{eq1a}) (D) Flow asymmetry $\varepsilon$ vs. initial leaflet angle $\theta_0$. The red line corresponds to a fit from Eq. (\ref{eq6}).}
	\label{fig1}
\end{figure}

These so-called fluid-structure interactions are observed in many biological systems \cite{doi:10.1152/ajprenal.2000.279.4.F698,doi:10.1152/ajprenal.1997.272.1.F132,Sotiropoulos2016} and have inspired a body of work characterizing the deformation of soft structures in flows \cite{wexler2013,Gomez2017,Jensen2018, Duprat2016,Amir5778,Moore2018,alvarado2017,Julien10612,Cappello2019}. The specificity of lymphatic valves lays in their unique shape and arrangement allowing for a very efficient way to passively create unidirectional flows. Yet, associating the valves shapes to the flow asymmetry and thus to their pumping efficiencies has been proven to be challenging and is usually limited to empirical models \cite{Moore2018,jamalian2017}. 

Here, in order to characterize how the soft valves induce flow asymmetry, we design a channel containing 3D printed pairs of valves inspired by 3D scans \cite{zawieja2009} (Fig. \ref{fig1} A). We first demonstrate experimentally the existence of an optimum valve geometry where the flow asymmetry is maximal. We then rationalize these results by performing experiments and simulations on a simplified geometry retaining only the essence of the fluid-structure interaction. This allows us to develop an analytical model that provides design guidelines to maximize the flow asymmetry and fluid pumping using soft valves.

This rationalization of the relationship between valve geometry and flow properties is a first step towards a better understanding of the malfunction of real lymphatic valves \cite{Moore2018,jamalian2017}. Furthermore, with the development of bio-printing \cite{Grigoryan458}, optimum valve geometries can now be weaved into living systems. Beyond biology, soft valves have recently inspired new active \cite{Unger113,Holmes1,Holmes2} and passive \cite{Gomez2017,alvarado2017} devices to control fluid flows in microfluidics as well as pressure driven soft robots based on fluid flows \cite{Vasios2019,wehner2016, Hines2017,Polygerinos2017,Rothemundeaar7986,Overvelde10863}. Therefore, a better characterization of soft valves inspired from the lymphatic system will allow to expend this mechanism to microfluidics and soft robotic systems where unidirectional flow could be induced not by the asymmetry of the actuation but by the asymmetry of the structure.

\section*{Lymphatic leaflets geometry}

Fig. \ref{fig1}A shows the schematics of 3D printed soft valves inspired by 3D scanners of lymphatic leaflets \cite{zawieja2009}. The valve is composed of two soft leaflets connected to the side of the channel. Similarly to lymphatic systems \cite{zawieja2009,Moore2018}, the valve is open at rest. Its aperture is defined by the initial angle $\theta_0$ and the length $a_{3D}$ (see Fig. \ref{fig1}A and Supplementary Information for further descriptions). The valve was printed out of elastic material (Young's modulus 1MPa, see methods) and scaled up by a factor $15$. To capture the influence of the valve on the flow under controlled conditions, we created a T-shaped channel, with one inlet and two outlets, where two valves are placed between the inlet and each outlet (see Fig. \ref{fig1}B). We filled the channel with silicone oil of viscosity 1 Pa.s, which guarantees a Reynolds number of $Re\le 10^{-1}$, comparable to Reynolds numbers involved in secondary lymphatic valves \cite{Moore2018}. We performed the experiments as follows: we impose the incoming flow rate $Q_{in}$ using a syringe pump and measure the flow rate of the two outlets $Q_{1,out}$ and $Q_{2,out}$. 

The measurements showed in Fig. \ref{fig1}C indicate a nonlinear inflow-outflow relation for both outlets. Moreover, the  outflow from outlet 1 $Q_{1,out}$ is higher than the outflow for outlet 2 $Q_{2,out}$ for any inflow $Q_{in}$. 
To quantify the asymmetry of the flow, we describe the non-linear pressure difference between inlet and outlet as
\begin{equation}
\Delta p=R Q(1+ \varepsilon Q),\label{eq1a}
\end{equation}
where $R$ and $\varepsilon$ are parameters depending on the valve geometry and orientation. In particular,  $\varepsilon$ describes the asymmetry of the flow. In this T-junction configuration, the only difference between the two lymphatic valves is their orientation. Therefore, we assume that the two resistances $R$ are equal and the asymmetries $\varepsilon$ only differ by their sign. Since the pressure differences on each side of the outlets are equal, we use Eq. (\ref{eq1a}) to derive analytical expressions of the outflows $Q_{1,out}$ and $Q_{2,out}$ as a function of $Q_{in}$ and fit them to the experimental data (Fig. \ref{fig1}C, red lines). The result characterizes well the asymmetry observed in Fig. \ref{fig1}C and allows to quantify the value of the asymmetry $\varepsilon$. From this set of experiments,  we provide evidences that a net flow is always induced in one specific direction with an intensity that is controlled by the inflow, which provides a robust mechanism for fluid pumping. Next, we show that this asymmetry can be optimized by the leaflets geometry.

We performed the same experiments for four pairs of valves with different initial angles $\theta_0=60^\circ, 70^\circ, 80^\circ,90^\circ$ and plotted in Fig. \ref{fig1}D the flow asymmetry $\varepsilon$ vs. the initial angle $\theta_0$. While very tilted $\theta_0=60^\circ$ or straight leaflets $\theta_0=90^\circ$ exhibit very small flow asymmetry, in between leaflet tilts ($\theta_0$=70$^\circ$ and 80$^\circ$), feature large flow asymmetry. This first result indicates the existence of a fine balance between the asymmetry of the leaflets and their interaction with the flow that allows for an optimum geometry maximizing flow asymmetry. However, the complex 3D geometry of the valve prevents the precise tracking of the valve deformation and therefore limits the conclusions and further studies on the geometric parameters governing the asymmetry. Therefore, in the remainder of the paper we focus on a idealized version of the lymphatic leaflets that have a 2D shape, which allows to easily quantify the leaflet deformation with the flow. This enables us to isolate and describe in details the precise interplay between asymmetry and nonlinearity and to develop an analytical model predicting the existence of an optimum leaflet geometry

\begin{figure}[b!]
	\centering
\includegraphics[scale=0.85]{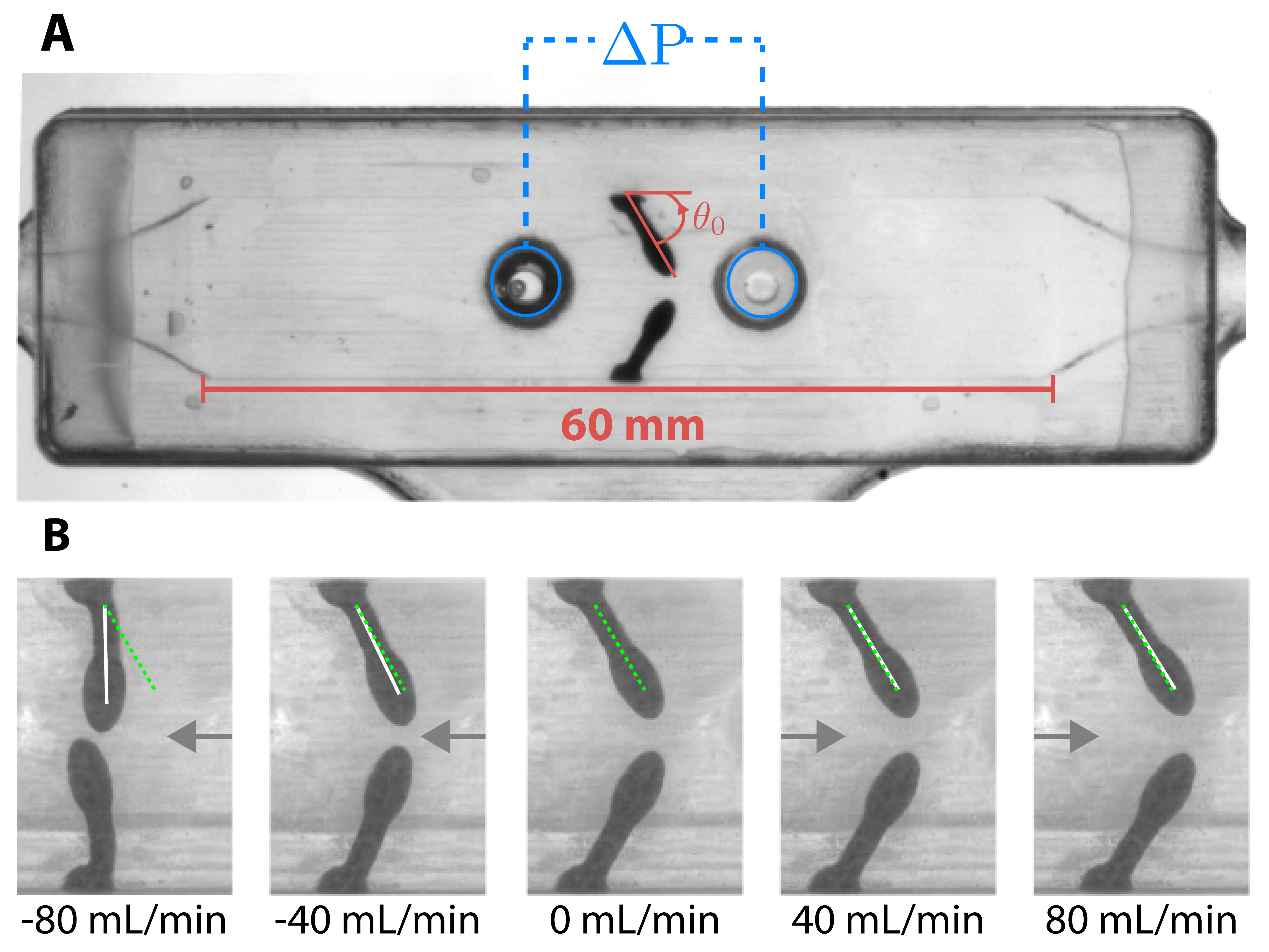}
			
	\caption{(A) A pair of 2D soft leaflets at an angle $\theta_0= 60^\circ$ in a rectangular rigid channel. The two holes highlighted in blue are connected to two pressure sensors. (B) Close up on the soft leaflets for different backward and forward flow rates.}
	\label{fig1b}
\end{figure}

\section*{2D leaflets geometry}

Using 3D printing, we create bio-inspired soft valves made of two soft leaflets placed on each side of a channel filled with water-glycerol mixture of viscosity $\eta=$ 0.3 $\pm 0.05$ Pa.s (Fig. \ref{fig1b}A). The two leaflets are made of a soft (Young's modulus $E_{leaflet}=2$ MPa) or a rigid (Young's modulus $E_{leaflet}=3$ GPa) material, while the rest of the channel is rigid (Young's modulus $E_{channel}=3$ GPa). The leaflets' height is $1.8$ mm, slightly smaller than the channel's height $2$ mm, such that they can freely bend. The pressure was measured at two points, on both sides of the leaflets. The new leaflets geometry allows to track their deformation with the flow (Fig. \ref{fig1b}B). While backward flows ($-80$ and $-40$ mL/min) largely reduce the gap for the liquid to flow, forward flows (40 and 80 mL/min) slightly increase the gap between both leaflets.

\begin{figure}[h!]
	\centering
\includegraphics[scale=0.85]{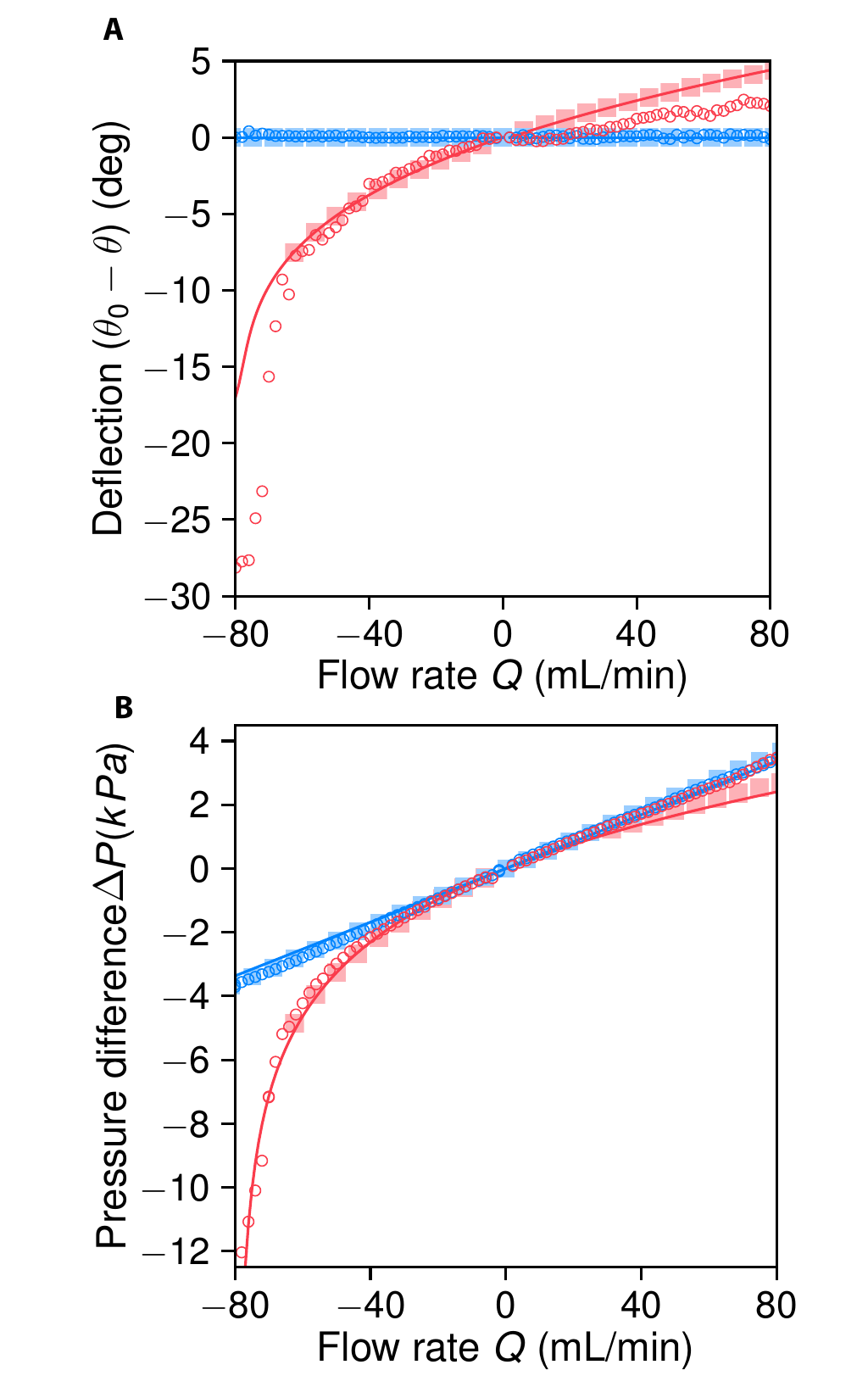}
			
	\caption{(A) Average deflection $\theta-\theta_0$ of rigid valves (in blue) and soft valves (in red) vs. flow rate $Q$. (B) Pressure difference $\Delta P$ measured on each side of the valve vs. flow rate $Q$ for rigid valves (in blue) and soft valves (in red). The points, squares and continuous lines correspond respectively to experiments, finite element simulations and the model described in Eq. (\ref{eq1}) and Eq. (\ref{eq2}).} 
	\label{fig2}
\end{figure}

We restrict our attention to flow rate $|Q|\le 80$ mL/min, which guarantees a Reynolds number $Re \le O(10^{-1})$ and therefore a laminar flow. We first measured the average deflection of the two leaflets as a function of the applied flow rate for both rigid and soft valves. As expected, the rigid leaflets (Fig. \ref{fig2}A, blue circles) do not deform with the flow. On the contrary, the soft leaflets deviate linearly from their initial angle $\theta_0$ (Fig. \ref{fig2}A, red circles). At high backward flows, the angle $\theta$ varies rapidly as the gap between the leaflets reduces. 
The deformation of the leaflets has a direct influence on the geometry of the channel, which in turn affects the pressure drop $\Delta P$ induced by the leaflets. For rigid leaflets, $\Delta P$ evolves linearly with the flow rate (Fig. \ref{fig2}B, blue circles), which is consistent with the fact that the flow is laminar and that the leaflets are immobile. In contrast, for soft leaflets the pressure $\Delta P$ evolves non-linearly with the flow rate $Q$ (Fig. \ref{fig2}B, red circles). As the flow rate increases towards positive (negative) values, the pressure difference increases (decreases) slower (faster) than the rigid valve case. In other words, the leaflets deformations do not only lead to non-linearities but also to an asymmetry in the pressure-flow relationship. This asymmetry is even more pronounced at high backward flows where the pressure diverges similarly to the deflection of the leaflets. This first series of observation confirm the conclusions from the experiments performed on lymphatic valves. In particular, contrary to an on-off valve, the soft leaflets induce asymmetry at any flow rate and for relatively small pressure difference.

\begin{figure}[t!]
	\centering
\includegraphics[scale=0.85]{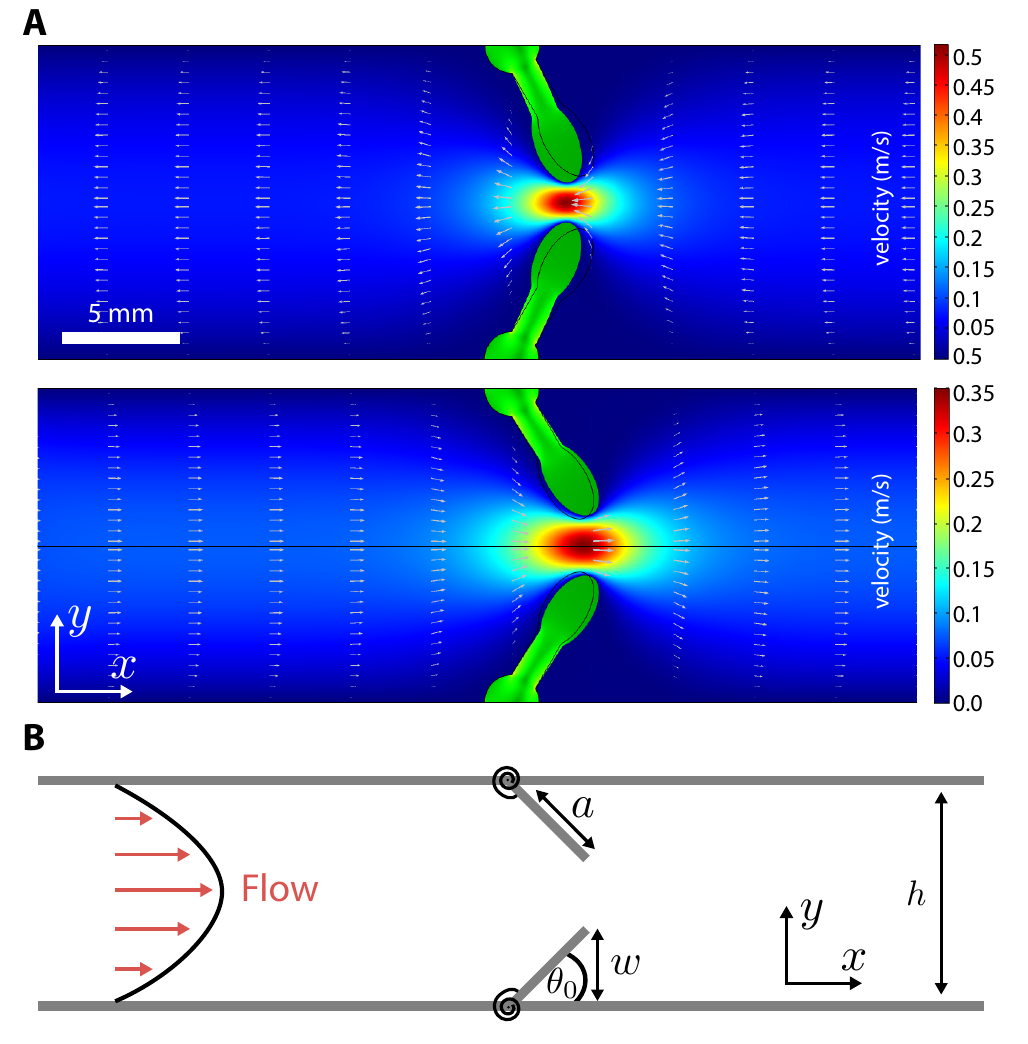}
			
	\caption{(A) 2D Simulation of the flow field around the soft leaflets (See Supplementary Information for a description of the pressure field). (B) Theoretical description of the experiment where the valves are sketched as two rigid plates connected to the sides of the channel by torsion springs (black spirals). The two leaflets and the channel have a depth $b$.}
	\label{fig3}
\end{figure}

To characterize the non-linear pressure-flow relationship, we simulate the flow and elastic deformations using a finite element 2D model (Comsol 5.4) considering a laminar flow and linear elastic approximation for the leaflets. The numerical simulation domain is a 2D channel of length $L= 40$ mm and width $h= 5.1$ mm that contains the 2D projections of one leaflet in its center (see Fig. \ref{fig3}A). The length $L$ is chosen long enough for the Poiseuille flow to fully develop before reaching the valve. The shape of the leaflet in the simulation is taken identical to the shape used in the experiment and we considered non-slip boundary conditions along the channel and the leaflets. To reduce calculation time, half of the system--below or above the central axis of the channel--is simulated by virtue of symmetry. The simulation box together with its mirrored image is shown in Fig. \ref{fig3}A for clarity. The simulation yields the deflection of the leaflet for different flow rates and is in good agreement with the measurements on both rigid and soft valves (see Fig. \ref{fig2}A, red and blue squares). By matching the pressure obtained from the simulations and the experiments performed on the rigid valve (Fig. \ref{fig2}B blue circles and squares) with a simple scaling factor, we also quantitatively capture the influence of the soft valve on the pressure flow relationship (Fig. \ref{fig2}B, red squares).
Therefore, the 2D finite element simulation captures very well the physics of the fluid-structure interaction, even if it does not take into account some 3D aspects of the experimental device such as the influence of the small gap between the leaflets and the top and bottom of the channel.

To capture the physics of the leaflets deformation within the flow, we model them as two infinitely thin and rigid plates of length $a$ and angular position $\theta$ connected to a torsion spring (black spirals in Fig. \ref{fig3}B), which is a reasonable assumption given the fact that elastic deformations of the leaflets are localized close to their anchoring points (see Fig. \ref{fig1b}B). Given that the 2D  finite element simulations capture well the experimental results, we model the flow around the leaflets as a 2D flow, therefore averaging the influence of the flow in the $z$ direction, and scale up the pressure by the same geometrical factor used in the numerical simulations. We consider the flow around the valve leaflet assuming low angles $\theta$ and use lubrication theory \cite{guyon2001hydrodynamique} to make the hypothesis of a parabolic flow profile between the two leaflets (see Fig. \ref{fig3}B and Supplementary Information). Therefore, the total pressure before the leaflets is expressed as
\begin{equation}
p= p_{end}+ \frac{12 \eta \beta q}{b } \int_{0}^{a \cos{(\theta)}} \frac{1}{(h- 2 x \tan{(\theta)})^3} dx',\label{eq1}
\end{equation}
where $p_{end}$ is the pressure after the valve, $q$ is the flow rate and $\eta$ is the liquid viscosity, $b$ is the channel's thickness and $\beta/b$ is the geometrical correction factor calibrated from the simulations.

The equilibrium angle of the leaflets $\theta$ is set by the torque balance between the torque exerted by the flow on the leaflet and the elastic restoring torque of the leaflet,
\begin{equation}
b \int_{0}^{a \cos{(\theta)}} x \delta p(x) dx = k (\theta- \theta_0) , \label{eq2}
\end{equation}
where $k$ is the torsional stiffness, $\delta p(x)= \frac{12 \eta \beta q}{b } \int_{x}^{a \cos{(\theta)}} \frac{1}{(h- 2 x \tan{(\theta)})^3} dx'$ is the pressure applied by the flow on the valve and $\theta_0$ is the angle in the reference state. Eqs.~(\ref{eq1}) and (\ref{eq2}) demonstrate the very nature of the nonlinear fluid structure interaction, where the flow deforms the structure (Eq. (\ref{eq2})), which in turn reconfigures the channel's geometry and affects the pressure difference (Eq. (\ref{eq1})). We non-dimensionalize Eqs.~(\ref{eq1}) and (\ref{eq2}) by introducing the variables
$\tilde{x} = x/L$, $\tilde{\Delta p(x)} = (p(x)-p_{end})/p_0$  and $\tilde{q} = q/q_0$, where $q_0= \frac{b h^3 p_0}{12 \beta \eta a}$ and $p_0=\frac{k}{ba^2}$. The new set of variables shows that the key parameters controlling the non-linearity are the torsional stiffness $k$, the viscosity $\eta$, the length ratio $h/a$ and the initial angle $\theta_0$. We numerically solve Eq. (\ref{eq2}) to deduce the position of the valve $\theta$ and from there inject $\theta$ in Eq. (\ref{eq1}). The stiffness $k= 33.5$ $\mu$N.m/rad was deduced from the valve Young's modulus, its length and second moment of area.
The results are shown in  Fig. \ref{fig2}AB and are in quantitative agreement with the experiments and the simulations despite the simplicity of the model.  This result shows that the physics of the leaflet deformation and its influence on the flow lays only in the $xy$ plane, which is the plane of symmetry of the soft valve. This means that the same approach could potentially be used for more complex 3D geometries. With such a predictive theoretical description at hand, we can now efficiently explore the relationship between geometry and flow properties.

\section*{Optimal geometry for maximal asymmetric flows}
\begin{figure}[t!]
	\centering
\includegraphics[scale=0.85]{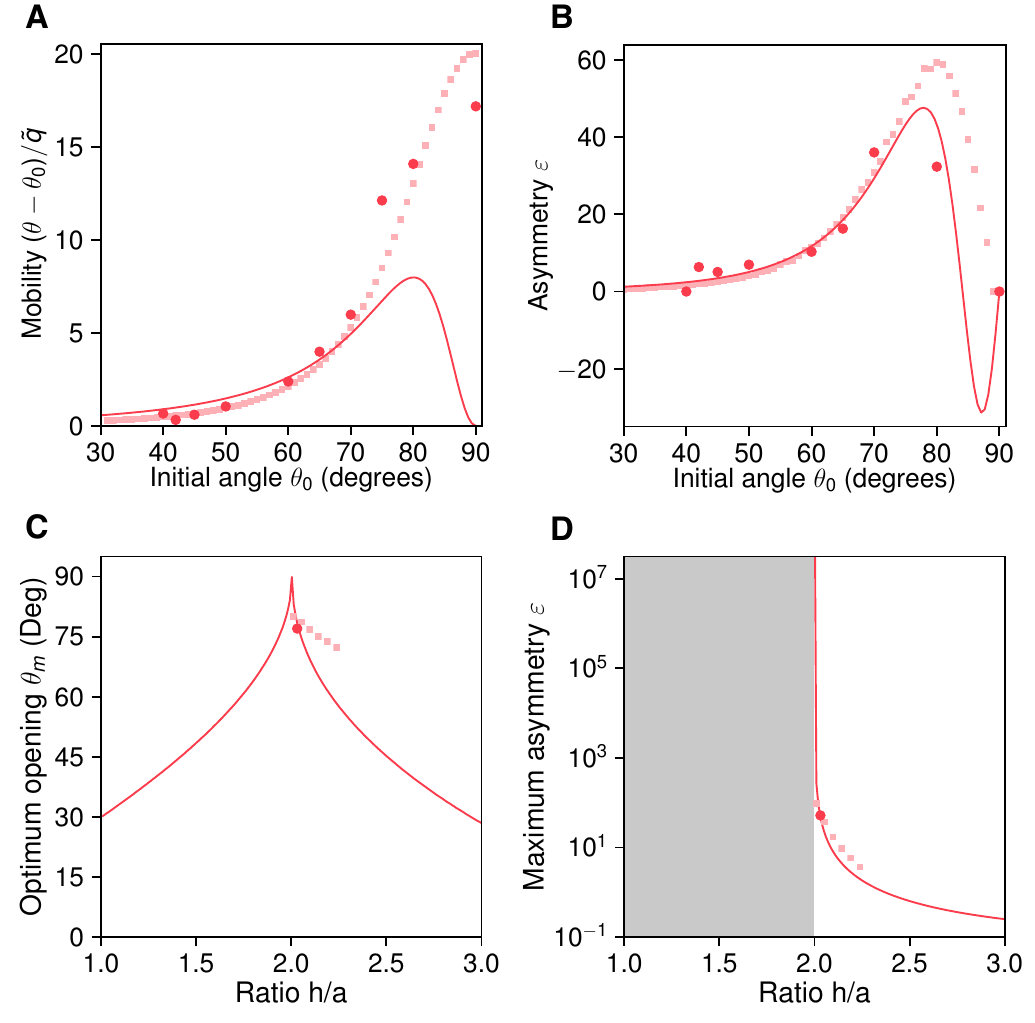}
			
	\caption{(A) Mobility of the leaflet $(\theta-\theta_0)/\tilde{q}$ vs. initial opening $\theta_0$. (B) Flow asymmetry $\varepsilon$ (see Eq. \ref{eq6}) vs. initial angle $\theta_0$.  (C) Initial opening $\theta_0^{\textrm{max}}$ at which the flow asymmetry $\varepsilon$ is maximized vs. geometric ratio $h/a$. (D) Corresponding maximum value of flow asymmetry vs. aspect ratio $h/a$. The continuous lines correspond to the model (see Eq. (\ref{eq4})). The red points (squares) are obtained from quadratic fits to the experimental results (simulations).}
	\label{fig4}
\end{figure}

Which leaflet geometry does induce a maximally nonlinear and asymmetric flow? To address this question, we solve Eqs.~(\ref{eq1}) and (\ref{eq2}) perturbatively, in the limit of  small deflections $\theta-\theta_0\ll 1$ and small flow rates $\tilde{q}\ll 1$. Following the observations from Fig. \ref{fig2}A, we assume that the deflection $\theta-\theta_0$ evolves linearly with the flow:
\begin{equation}
\theta-\theta_0=f(\alpha_0) \cos^2(\theta_0)  \tilde{q},\label{eq32}
\end{equation}
where $f(\alpha_0)$ is a function of $\alpha_0=2 \sin(\theta_0) a/h$ (see Supplementary Information). The linear approximation is consistent with the observation from Fig. \ref{fig2}A at small deflections. We expect the angular slope $(\theta-\theta_0)/ \tilde{q}$ to increase monotonically with the initial angle $\theta_0$. Indeed, as the leaflets become perpendicular to the flow, the lever arm of the flow on the leaflets, hence its torque, increases. This is confirmed by the experiments and the simulations in Fig. \ref{fig4}A (red points and squares). Remarkably, the analytic prediction is in quantitative agreement with the numerical and experimental results up to high initial openings $\theta_0$$>$70$^\circ$ (Fig. \ref{fig4}A, red line). Beyond this angle, the lubrication theory and the assumption of infinitely thin leaflets break down. With an analytical expression of the deflection $\theta-\theta_0$, the difference in pressure between the leaflets as a function of the flow rate can be rewritten at the quadratic order as (See Supplementary Information)
\begin{equation}
\tilde{\Delta p}=R \tilde{q}(1+ \varepsilon \tilde{q}),\label{eq4}
\end{equation}
where $R$ and $\varepsilon$ depend on the geometric parameters $\theta_0$, $a/h$ and $\alpha_0$
\begin{equation}
R=\frac{\cos{(\theta_0)}}{2} \frac{(2-\alpha_0)}{ (1-\alpha_0)^2},\label{eq5}
\end{equation}
and
\begin{equation}
\varepsilon=f(\alpha_0) \left ( \frac{  a (3-\alpha_0)\cos^3(\theta_0)}{h (2-\alpha_0)(1-\alpha_0)}-\frac{ \sin{(\theta_0)} \cos{(\theta_0)}}{2}. \right)\label{eq6}
\end{equation}
The pre-factor $\varepsilon$ is exactly the same as the one discussed for the lymphatic leaflet geometry and has now an analytical expression deduced from the 2D leaflet geometry. The Eq. (\ref{eq6}) quantifies the asymmetry of the pressure-flow relationship observed in Fig. \ref{fig2}B. The experiments and finite element simulations performed on the 2D leaflet geometry and Eq. (\ref{eq6}) all show that this asymmetry $\varepsilon$ varies non-monotonically with the initial opening $\theta_0$ (see Fig. \ref{fig4}B). For small initial angles $\theta_0$, the valves are barely deformed by the liquid flow (Fig. \ref{fig4}A), leading to a linear pressure flow-rate relationship and small flow asymmetry $\varepsilon$. As the initial angle increases, the valve is more deflected by the flow and the asymmetry between forward and backward flows increases. In contrast, a symmetric channel with an initial angle $\theta_0 = 90^\circ$ naturally has no flow asymmetry $\varepsilon=0$. 
This competition between geometric asymmetry and lever arm effects explains the observation of an optimum initial angle at which the asymmetry between forward and backward flows is maximized. In Fig. \ref{fig4}B, this maximum is reached for $\theta_0= 78^\circ\pm3^\circ$. 

How does the optimal angle $\theta_0^{\textrm{max}}$ and the corresponding flow asymmetry depend on the aspect ratio between the leaflet's length $a$ and the channel width $h$? To answer this question, we use our analytical results described above and run additional numerical simulations where we vary the aspect ratio $h/a$ (Fig. \ref{fig4}C). For aspect ratios $h/a < 2$, i.e. leaflets longer than half the channel size, the maximal asymmetry is always obtained for leaflets that are initially touching, hence acting as an on-off valve. In this case, the initial angle $\theta_0^{\textrm{max}}$ maximizing the flow asymmetry can be simply derived from a geometric relation $\theta_0^{\textrm{max}}=\arcsin(h/(2a))$, which therefore decreases for smaller aspect ratios. In order words, the longer the leaflet, the more it should be tilted to induce an optimal asymmetry and nonlinear flow (Fig. \ref{fig4}C). In contrast, for aspect ratios $h/a > 2$, the geometrical balance is more intricate, and we find that for shorter leaflets, optimally nonlinear and asymmetric flows are obtained for more tilted leaflets (Fig. \ref{fig4}C). This can be intuitively understood from the fact that for large ratios $h/a$, the leaflets are less confined. Therefore, they influence less the flow around $\theta_0 \approx 90^\circ$ and the maximum observed in Fig. \ref{fig4}B shifts towards lower angles $\theta_0$.

We calculated the level of asymmetry at the optimum angle for different ratios $h/a$ (Fig. \ref{fig4}D). For $h/a<2$, any backward flow induces self-contact of the leaflets, triggering an on-off valve effect and leading to an infinite pressure and a theoretically infinite flow asymmetry $\varepsilon$. For $h/a>2$ no such contact occur, the pressure difference remains finite, and exponentially decreases with $h/a$. While it appears that maximum levels of asymmetry are reached for $h/a<2$, note that the resistances $R$ described by Eq. (\ref{eq5}) also diverges at $h/a=2$ (see Supplementary Information). Ratios $h/a>2$ can therefore be optimum for systems that require small resistance $R$ to generate flow. 

To further validate these analytical predictions, we used our analytical findings to describe the measurements performed on the leaflets with the lymphatic valve geometry. We calibrated the minimum length $a_{3D}$ in the center of the 3D-leaflets (Fig. \ref{fig1}A) and fitted with Eq. (\ref{eq6}) the asymmetry measured in Fig. \ref{fig1}C with the torsional stiffness $k$ as the only free parameter. The fit is in excellent agreement with the measurements and gives a torsional stiffness of the same order of magnitude as the 2D valve case. This result demonstrates the generality of our approach and its applicability to complex shapes. Furthermore, it rationalizes the in-vivo observations of lymphatic valves opened at rest \cite{Moore2018}.  

To conclude, we have used bio-inspired soft leaflets to evidence a generic mechanism, where a nonlinear interaction between a laminar flow and a flexible structure induces a nonlinear and asymmetric flow response. Our findings provide vistas to quantitatively predict such asymmetric flow response, rationalize the shape and mechanical properties of leaflets in living organisms and create designer channels with optimal flow asymmetry. In particular, we establish design rules to maximize the flow asymmetry with the leaflet geometry. We envision that our results provide a step forward to gain a better understanding of the role of the lymphatic valves in the lymphatic system \cite{Moore2018,jamalian2017}, where accurate measurements of the valve geometry for specific specimens would allow to predict their efficiency. This research also opens novel avenues for bio-inspired microfluidic \cite{Unger113,Holmes1,Holmes2} and soft robotic  \cite{Vasios2019,wehner2016, Hines2017,Polygerinos2017,Rothemundeaar7986,Overvelde10863} devices, where internal fluid-structure interactions could passively control flows.

{\em Acknowledgements}  We are grateful to T. Gilet and  A. Schiphorst for insightful discussions. We thank D. Giesen, S. Koot and G. Hardeman for skillful technical support, L. D. Thanh, A. Schiphorst, E. Peerbooms and S. Broersen for preliminary experiments. We acknowledge funding from the Netherlands Organization for Scientific Research (NWO)  via a VENI grant.

\section{Supporting Information}
\setcounter{figure}{0}
\renewcommand{\figurename}{Figure}
\renewcommand\thefigure{S\arabic{figure}}
\setcounter{equation}{0}
\renewcommand\theequation{S\arabic{equation}}

\subsection{Experimental method}

\begin{figure}[b!]
	\centering
\includegraphics[scale=0.6]{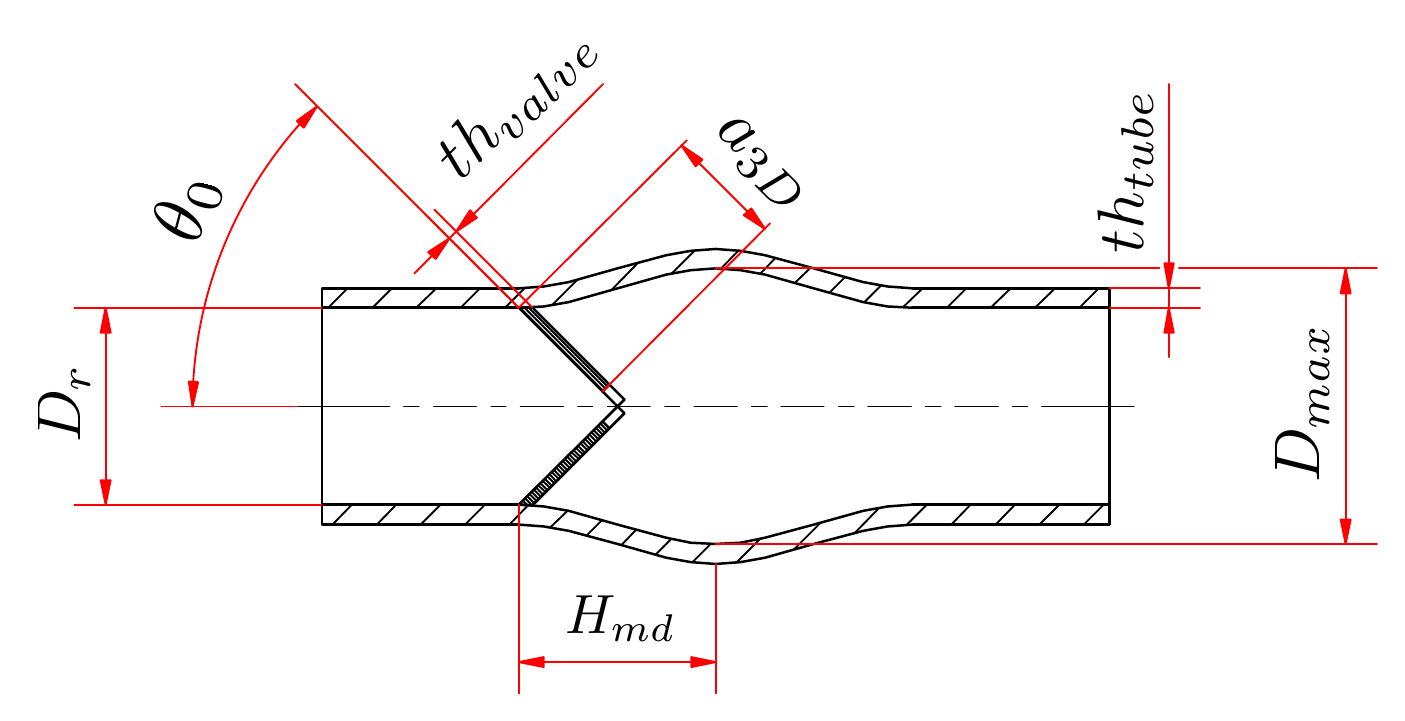}
			
	\caption{Longitudinal cut of the schematic inspired from 3D scanners of lymphatic leaflets\cite{zawieja2009}.}
	\label{SI1}
\end{figure}

\subsubsection{3D leaflets}

The design of a 3D printed channel is shown in Fig. 1A of the Main Text and in Fig. \ref{SI1}. The cylindrical channel has a thickness $th_{tube} = 1$ mm and a minimum internal diameter $ D_r = 10$ mm. The enlargement in the middle, corresponding to the sinus region, has a larger radius $D_{max} = 14 $ mm reached over a distance $H_{md}= 10$ mm. The 3D leaflet is connected to the cylinder on its side. The initial angle $\theta_0$ is measured at the center of the leaflet and was varied from 60$^\circ$ to 90$^\circ$. The leaflet minimum length $a_{3D} = 4.9$ mm and its thickness  $th_{valve}=1$ mm were kept constant. The whole valve was printed out of \textit{Agilus 30 Clear} (Young's modulus 0.6 MPa). We injected silicone oil of viscosity 1 Pa.s using a syringe pump (Harvard apparatus PhD Ultra) at a controlled flow rate $Q_{in}$ to study the fluid-structure interaction. We measured the two outflows $Q_{1,out}$ and $Q_{2,out}$ by determine the mass rate of the silicone oil coming out of both ends.

\subsubsection{2D leaflets}

The 2D bio-inspired setup is shown in Fig. 2A of the Main Text. The main channel is a rectangle of 60x10 mm with a depth of 2 mm. The left and right sides of the rectangular channel are connected to tubes that are themselves connected to a syringe pump (Harvard apparatus PhD Ultra)  in push/pull configuration, which allows to work in a closed environment. The channel is filled with a water-glycerol mixture at a viscosity of $\eta=$0.3 $\pm 0.05$ Pa.s. The two holes inside the rectangular channel (highlighted by blue circles) are connected to two pressure transducers (TR series TR1-0015G-101) transmitting the pressure measurements to a computer via a Data Acquisition System (LabJack U6-PRO). The valve leaflets situated in the center of the design were printed together with the channel by using a multi-material 3D printer (Stratasys Objet500 Connex3). The channel was made out of \textit{Fullcure 720} (Young's modulus 3 GPa). The same material was used to print rigid leaflets, while soft leaflets were printed out of a mix of \textit{Agilus 30 Clear} and \textit{Veroblack plus} (Young's modulus 2 MPa). Once printed, a transparent plexiglass plate of thickness 5 mm was glued on top the rectangular channel to seal it. The shape of the valve at a given flow rate was recorded using a camera Basler acA3800-14um (3840 px x 2748 px resolution) with the Basler Lens C125-1620-5M F2.0 f16mm (2742 px/mm).

Fig. 2B of the Main Text shows a close up on the leaflets. Each valve has the same shape. The rounded base, connected to the rigid channel ensures the cohesion between both printed materials.  The slightly thicker tip of the valve was designed to avoid valve twisting along its axis while allowing bending in the x-y plane. The valve has a length $h=5.12$ mm and a minimum (maximum) width $r_{min}=1$ mm ($r_{max}= 1.5$ mm). The thickness of the valves was kept slightly smaller than the size of the channel (1.8 mm) to prevent friction while avoiding liquid to flow around the valves.

\subsection{Numerical simulations} \label{simu}

Numerical simulations have been performed considering a 2D projection of the experiment and assuming an incompressible and stationary flow at low Reynolds numbers. Therefore, we solved the following Stokes equations in 2D
\begin{equation}
 \frac{\partial  p}{\partial x}+\eta (\frac{\partial^2v_x}{\partial x^2}+\frac{\partial^2v_x}{\partial y^2})=0\label{eq-2},
\end{equation}
\begin{equation}
\frac{\partial p}{\partial y}+\eta (\frac{\partial^2v_y}{\partial x^2}+\frac{\partial^2v_y}{\partial y^2})=0\label{eq-1},
\end{equation}
where $\eta$ is the viscosity of the fluid and $p$ is the pressure within the flow.
\begin{figure}[t!]
	\centering
\includegraphics[scale=0.8]{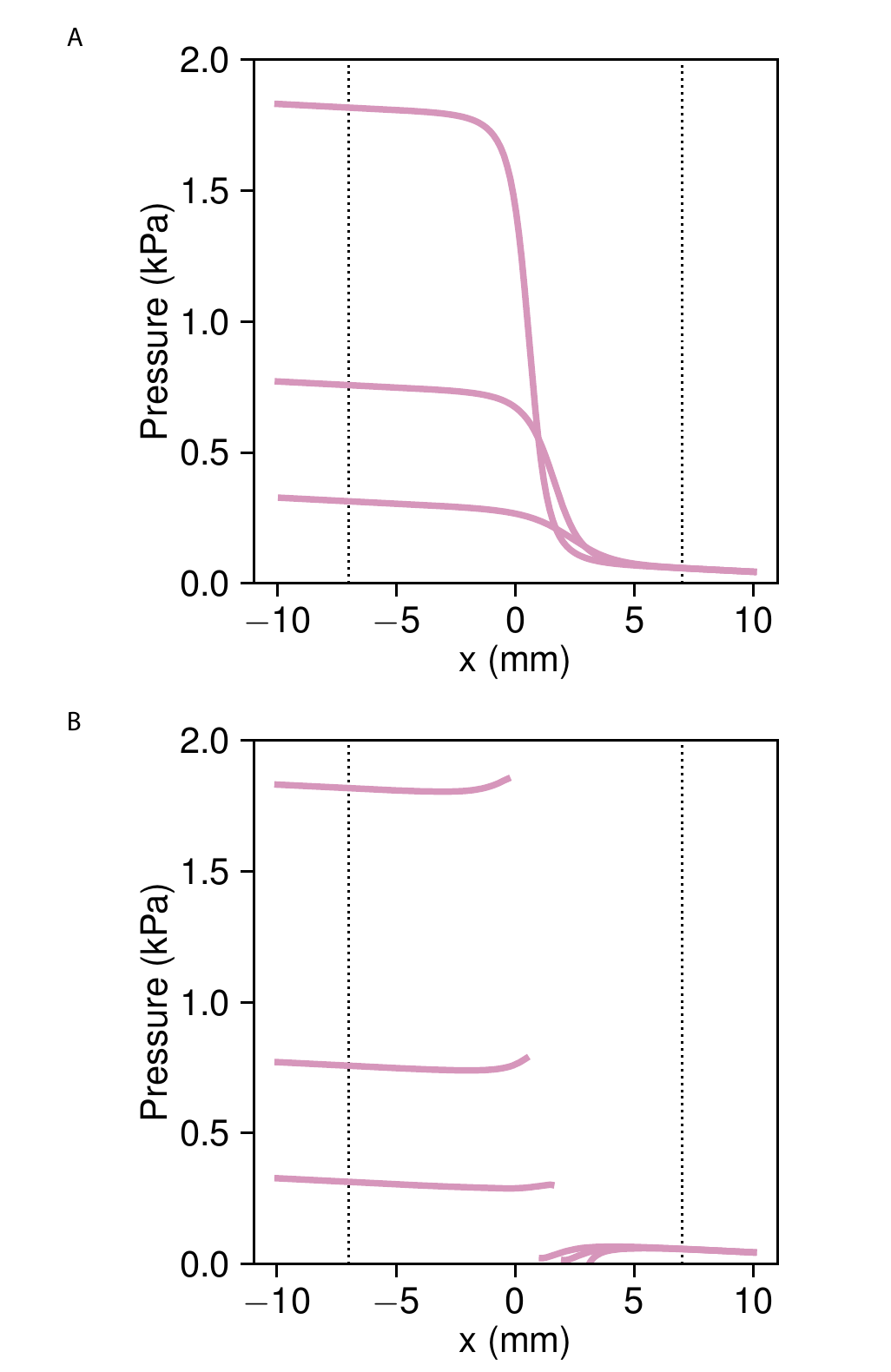}
			
	\caption{Pressure in the middle of the channel (A) or in-between the middle of the channel and its base (B) issued from numerical simulations for positive flow rates and three initial angles $\theta_0= $ 40$^\circ$,  60$^\circ$, 80$\circ$. The higher pressure corresponds to larger initial angles. The discontinuity in panel B corresponds to the position of the leaflet.}
	\label{SI3}
\end{figure}

The results of the simulations are shown in Fig. 3 and 4 of the Main Text. To further characterize the flow around both leaflets and confirm our theoretical assumptions, we plot in Fig. \ref{SI3} the pressure profile in the middle of the channel (Fig. \ref{SI3}A) and in-between the side of the channel and its middle (Fig. \ref{SI3}B) for 3 initial angles $\theta_0=$ 40$^\circ$, 60$^\circ$ and 80$^\circ$. The discontinuity in Fig. \ref{SI3}B is due to the presence of the leaflet. In both cases and for all 3 initial angles, the pressure quickly drops after the leaflet, indicating that the influence of the leaflet on the pressure is much more important than the pressure variation along the channel without leaflets. The vertical dashed lines describe the position at which the difference in pressure was measured both in numerical simulations and experiments. Fig. \ref{SI3} shows that the pressure barely drops before reaching the leaflet and after the leaflet, indicating that the measured pressure difference captures well the difference in pressure on each side of the valve. Moreover, by comparing Fig. \ref{SI3}A and Fig. \ref{SI3}B, we observe that the pressure barely varies along the $y$ direction.

\begin{table}[ht]
 \begin{center}   {
\begin{tabular}{@{} l L L L  @{} >{\kern\tabcolsep}c @{}} 

 \toprule
  & 40$^\circ$ & 60$^\circ$ & 80$^\circ$         \\    \midrule
 $\partial_{yy} v_x$ &  0.08479  &  0.29510 &  1.09355     \\   
\rowcolor{black!20}[0pt][0pt]    $\partial_{xx} v_x$ & 0.00033 &  0.00023 &  0.00322    \\   
 $\partial_{yy} v_y$ &  0.00108  &  0.07872 &  0.14103     \\   
\rowcolor{black!20}[0pt][0pt]    $\partial_{xx} v_y$ & 9.12 10$^{-14}$ &  9.67 10$^{-14}$&  5.20 10$^{-14}$    \\   
 $\partial_x p$ &  0.31708 &  0.76146 & 1.82213     \\  
 \rowcolor{black!20}[0pt][0pt]    $\partial_y p$ & 0.02373 &  0.05160  & 0.04604    \\  
   \bottomrule
  \hline

\end{tabular}}
 \end{center}
\label{Chap1tab1} 
\caption{\label{Chap1tab1} Orders of magnitude of different terms in Eq. (\ref{eq-2}) and (\ref{eq-1}) obtained from numerical simulation in-between the leaflets ($x=$ 1 mm and $y=$ 5.12 mm)}
\end{table}

\subsection{Theoretical model}

\subsubsection{Pressure-flow relationship}

Similarly to the numerical simulations, we first consider the flow in-between the leaflets as a 2D problem. By assuming a stationary and incompressible flow, the $x$ and $y$ components of the speed are described by Eqs. (\ref{eq-2}) and (\ref{eq-1}). We then use the lubrication theory and assume that the speed and pressure variation along the $y$ axis, $v_y$ and $\frac{dp}{dy}$ are negligible compared to the speed and pressure variation along the $x$ axis, $v_x$ and $\frac{dp}{dx}$. 
\begin{equation}
 0= \frac{\partial  p}{\partial x}+\eta \frac{\partial^2v_x}{\partial y^2}\label{eq-2bis}.
\end{equation}
This assumption is expected to work only for small variation in width  $h-w(x)$, where $h$ is the width of the channel and $w(x)$ the deflection of the leaflet at a point $x$ (i.e. for small $\theta$, see Fig. 4B of the Main Text). However, numerical simulations examining the flow speed and its gradients in between both leaflets (at x=1mm and y= 5.12 mm in Tab. 1) show that the hypothesis stays valid for larger angles $\theta$. With the lubrication assumption, we obtain:
\begin{equation}
v_x(y)= - \frac{dp}{dx} \frac{h-2w(x)}{2 \eta} y (1-\frac{y}{h-2w(x)})\label{eq0},
\end{equation}
where $h-2w(x)$ is the distance between both leaflets and decreases as $x$ increases. From the expression of the speed, we deduce the $2D$ flow rate
\begin{equation}
Q_{2D}= - \frac{dp}{dx} \frac{ (h-2w(x))^3}{12 \eta}\label{eq0bis},
\end{equation}
we now focus on relating this $2D$ flow rate to the flow rate measured in the 3D channel. Since the $z$ direction has no impact on the fluid structure interaction, we average the flow gradient in the $z$ direction by multiplying Eq. (\ref{eq0bis}) by $b/\beta$, where $b$ is the height of the channel and $\beta$ is a geometrical factor calibrated on measurements performed with rigid valves, where the configuration of the channel does not vary with the flow rate. Therefore, we obtain
\begin{equation}
Q= - \frac{dp}{dx} \frac{b (h-2w(x))^3}{12 \beta \eta}\label{SIeq1},
\end{equation}
The simulated flow profile showed in Fig. \ref{SI3} shows that the flow field comes back very quickly to a Poiseuille flow inside a channel of size $h$ x $b$ after the valve. Therefore, we assume that the flow rate after the valve is

\begin{equation}
Q= - \frac{dp}{dx} \frac{b h^3}{12 \beta \eta}\label{SIeq2},
\end{equation}
integrating the two previous equations, the pressure upstream of the leaflets is

\begin{equation}
p=p_{end}+ \frac{12 \beta \eta Q}{b } \int_{0}^{\ell} \frac{1}{(h- 2 w(x'))^3} dx'\label{SIeq3},
\end{equation}
where $\ell$ is the length of the leaflets projected on the $x$ axis and depends on the position of the leaflets. The pressure at the downstream end of the valve $p_{end}$, measured relative to the ambient pressure is given by

\begin{equation}
p_{end}=\frac{12 \beta \eta Q L}{b h^3},
\end{equation}
where $L$ is the length of the channel at the downstream end of the valve. In practice, $L \gg \ell$ such as $L$ does not depend on the leaflets position. Given the small variation in pressure before the leaflets and after the leaflets observed in the numerical simulations (see Fig. \ref{SI3}), we assume that the pressure sensors exactly measure the pressure difference on each side of the valve $p-p_{end}$, even if they are situated few centimeters away from the valve.

\begin{figure}[t!]
	\centering
\includegraphics[scale=0.7]{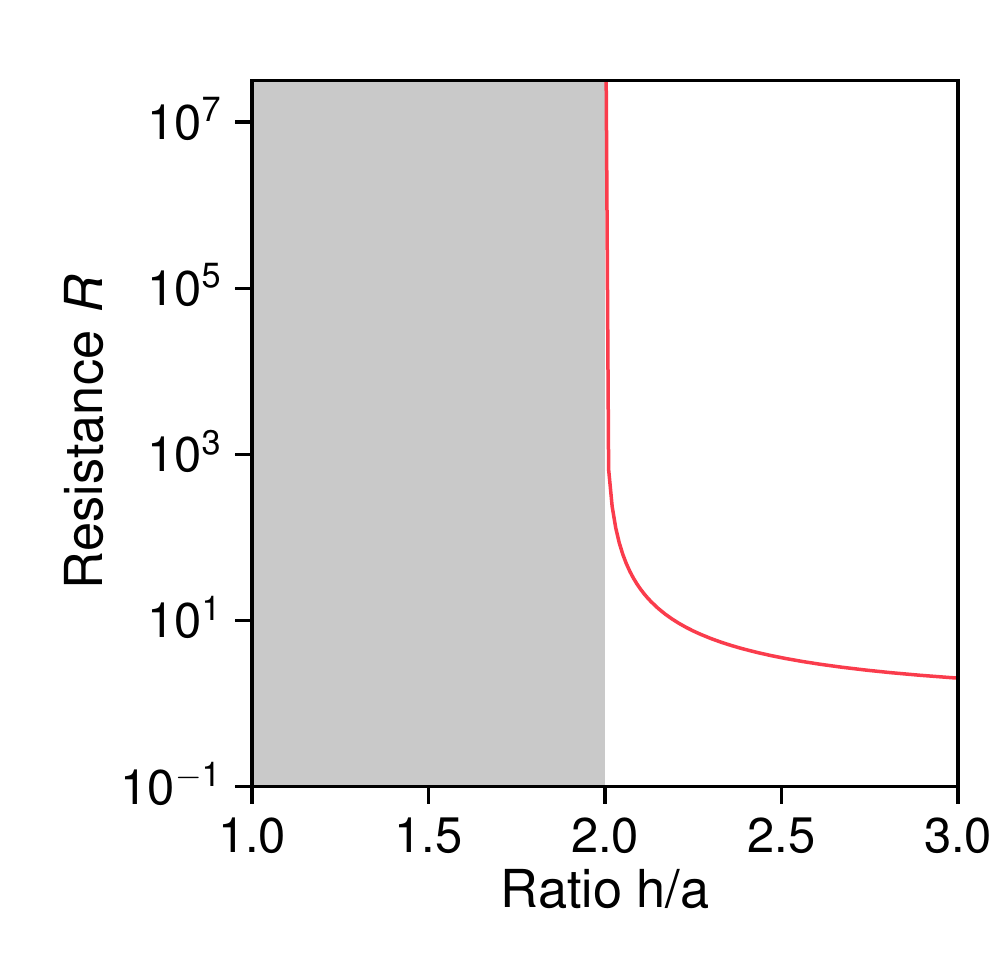}
			
	\caption{Resistance $R$ deduced from Eq. (6) of the Main Text as a function of $h/a$.}
	\label{SI4}
\end{figure}

The next step to relate the pressure difference to the flow rate is to characterize the deformation of the leaflets. To this end, we model the leaflets as an infinitely thin plate of length $a$ (see Fig. 4B of the Main Text). Therefore, we have $w(x) = x \tan{(\theta)}$ and $l =a \cos{(\theta)}$, where $\theta$ is the angular position of the leaflets.  The stiffness of the leaflets is modelled as a torsion spring of stiffness $k$ located at the base of the rigid plate. The equilibrium angle of the leaflets $\theta$ is set by the balance between the torque applied by the flow on the rigid plate and the torsion spring
\begin{equation}
\int_{0}^{a \cos{(\theta)}} x' \delta p(x') dx' = k (\theta- \theta_0) \label{SIeq5},
\end{equation}
where $\delta p(x')=\frac{12 \beta \eta Q}{b } \int_{x'}^{l} \frac{1}{(h- 2 w(x''))^3} dx''$ is the difference in pressure felt by the leaflets at a point $x'$. We assume that the pressure difference $\delta p(x')$ is only dictated by the flow in-between the leaflets,which is a reasonable assumption given the fact that the pressure below the leaflet is approximately equal to the pressure measured after the leaflet (Fig. \ref{SI3}B). The parameter $\theta_0$ is the initial angle of the leaflets.  We non-dimensionalize Eqs. (\ref{SIeq3}) and (\ref{SIeq5}) by introducing the variables $\tilde{x} = x/L$, $\tilde{\Delta p} = (p-p_{end})/p_0$  and $\tilde{q} = q/q_0$, where $q_0= \frac{b h^3 p_0}{12 \beta \eta a}$ and $p_0=\frac{k}{ba^2}$. Therefore, Eq. (\ref{SIeq5}) becomes
\begin{equation}
\frac{\tilde{q} \cos(\theta)^2}{4 \alpha^3 } \left ( \frac{\alpha (3 (\alpha-2)}{(\alpha-1)^2}- 2 \log(1-\alpha) \right )=\theta-\theta_0 \label{SIeq5bis},
\end{equation}
where $\alpha= 2 a \sin(\theta)/h$ . We numerically solve Eq. (\ref{SIeq5bis}) and obtain the angule of the leaflets $\theta$. From there, we calculate the pressure difference between both leaflets by inserting $\theta$ in 
\begin{equation}
\tilde{\Delta p}=  \frac{ \tilde{q}}{2}\frac{(2-\alpha)}{(1-\alpha)^2} \cos(\theta)\label{SIeq5bisbis},
\end{equation}
The resulting angular deflection $(\theta-\theta_0)$ and pressure are shown in Fig. 3 of the Main Text for an initial angle $\theta_0=60^\circ$.

\subsubsection{Analytical solution at low flow rates}
We further study the influence of the leaflet geometry on the pressure-flow relationship by expanding Eqs. (\ref{SIeq3}) and (\ref{SIeq5}) in the limit of small deflections $(\theta-\theta_0)$ and small flow rates $\tilde{q}$ and assume that $(\theta-\theta_0)$ varies linearly with  $\tilde{q}$, which is confirmed by the measurements and simulations in Fig. 4A of the Main Text. This leads to Eq. (4) of the Main Text, with
\begin{equation}
f(\alpha_0)= \frac{\alpha_0\frac{(3 \alpha_0-2)}{(\alpha_0-1)^2}-2 \log{(1-\alpha_0)}}{4 \alpha_0^3} \label{SIeq6},
\end{equation}
By expanding Eq. (\ref{SIeq5bisbis}) in the limit of small deflections $(\theta-\theta_0)$, we then obtain Eq. (5) of the Main Text. In Fig. \ref{SI4}, we show that $R$ behaves similarly as $\varepsilon$ (see Fig 5D of the Main Text) as a function of $h/a$. Therefore, while the asymmetry is optimized for $h/a < 2$, a higher pressure is required for the liquid to low both forward and backward. As a result, ratio $h/a>2$ is more optimum to generate flows without the need of high pressures.

\subsubsection{Analytical solution for a T-junction}

We seek for an analytical expression of the outflows $Q_{1,out}$ and $Q_{2,out}$ as a function of $Q_{in}$ measured in Fig. 1C of the Main Text. Knowing that the pressure on each side of the T-junction is equivalent and assuming that the pressure varies quadratically with the flow, we have

\begin{equation}
R_1 \tilde{q}(1+ \varepsilon_1 \tilde{q})=  R_2 \tilde{q}(1+ \varepsilon_2 \tilde{q}) \label{SIeq7},
\end{equation}
where $R_1$ ($R_2$) and $\varepsilon_1$ ($\varepsilon_2$)  correspond to the linear and asymmetric term of the left (right) side of the T-junction. Given the relative symmetry of the two sides, we assume $R_1=R_2$ and $\varepsilon_1=-\varepsilon_2=\varepsilon$. Therefore we obtain
\begin{equation}
Q_{out,1} =\frac{Q_{in}}{2}+\frac{\varepsilon Q_{in}}{4}\label{SIeq8},
\end{equation}

\begin{equation}
Q_{out,2} =\frac{Q_{in}}{2}-\frac{\varepsilon Q_{in}}{4}\label{SIeq9},
\end{equation}

These two functions have been fitted on measurements in Fig. 1C of the Main Text, which allows to quantify the asymmetry $\varepsilon$. Fig. 1D shows the asymmetry $\varepsilon$ has a function of the initial angle.





%

\end{document}